\documentclass[pra,reprint,twocolumn,showpacs,superscriptaddress]{revtex4-1}

\usepackage{graphicx}
\usepackage{dcolumn}
\usepackage{bm}
\usepackage{graphicx}
\usepackage{epsfig}
\usepackage{epsf}
\usepackage{amssymb}
\usepackage{amsmath,empheq,cases}
\usepackage{amsthm}
\usepackage{mathtools}
\usepackage{multirow}
\usepackage[colorlinks=true,linkcolor=blue,citecolor=blue,pdfauthor={ },pdftitle={ },pdfsubject={ },pdfkeywords={ }]{hyperref}
\usepackage{pgfplots}
\usepackage{lipsum}

\usepackage{times}

\theoremstyle{definition}

\usepackage{multirow}
\usepackage{dcolumn}
\newcolumntype{d}[1]{D{.}{.}{#1}}

\begin{document}

\title{Optimal Bounds on      State Transfer Under Quantum Channels with Application to    Spin System  Engineering}

\author{Wenqiang Zheng}
\affiliation{Collaborative Innovation Center for Bio-Med Physics Information Technology, College of Science, Zhejiang University of Technology, Hangzhou 310023, China}

\author{Hengyan Wang}
\affiliation{Department of Physics,  Zhejiang University of Science and Technology, Hangzhou 310023, China}

\author{Tao Xin}
\affiliation{Shenzhen Institute for Quantum Science and Engineering and Department of Physics, Southern University of Science and Technology, Shenzhen 518055, China}
\affiliation{Center for Quantum Computing, Peng Cheng Laboratory, Shenzhen 518055, China}
\affiliation{Shenzhen Key Laboratory of Quantum Science and Engineering, Southern University of Science and Technology, Shenzhen 518055, China}

\author{Xinfang Nie}
\email{niexf@sustech.edu.cn}
\affiliation{Shenzhen Institute for Quantum Science and Engineering and Department of Physics, Southern University of Science and Technology, Shenzhen 518055, China}

\author{Dawei Lu}
\email{ludw@sustech.edu.cn}
\affiliation{Shenzhen Institute for Quantum Science and Engineering and Department of Physics, Southern University of Science and Technology, Shenzhen 518055, China}
\affiliation{Center for Quantum Computing, Peng Cheng Laboratory, Shenzhen 518055, China}
\affiliation{Shenzhen Key Laboratory of Quantum Science and Engineering, Southern University of Science and Technology, Shenzhen 518055, China}

\author{Jun Li}
\email{lij3@sustech.edu.cn}
\affiliation{Shenzhen Institute for Quantum Science and Engineering and Department of Physics, Southern University of Science and Technology, Shenzhen 518055, China}
\affiliation{Center for Quantum Computing, Peng Cheng Laboratory, Shenzhen 518055, China}
\affiliation{Shenzhen Key Laboratory of Quantum Science and Engineering, Southern University of Science and Technology, Shenzhen 518055, China}

\begin{abstract}

Modern applications of quantum control in quantum information science and technology require the precise characterization of quantum states and quantum channels. In particular, high-performance quantum state engineering  often demands that quantum states are transferred with optimal efficiency via    realizable controlled evolution, the latter often modeled by quantum channels. When an appropriate  quantum control model for an interested system   is constructed,   the exploration of  optimal bounds on state transfer   for the underlying quantum channel  is then an important   task. In this work, we analyze the state transfer efficiency problem for different class of quantum channels, including unitary, mixed unitary and Markovian. We then apply the theory   to     nuclear magnetic resonance  (NMR)  experiments. We show that two most commonly used control techniques in NMR, namely gradient field control and phase cycling, can be described by mixed unitary channels. Then we show that  employing mixed unitary channels does not extend the unitarily accessible region of states. Also,   we present a strategy of optimal  experiment design,  which incorporates coherent radio-frequency field control, gradient field control and phase cycling, aiming at maximizing state transfer efficiency and meanwhile minimizing the number of experiments required. Finally, we perform pseudopure state preparation experiments on two- and three-spin systems, in order to test the bound theory and to demonstrate the usefulness of non-unitary control means.  
\end{abstract}


\maketitle

\section{Introduction}

Recent years have seen the extensive use of control theory in the active manipulation   of quantum systems \cite{DP10,BCR10}. 
The  subject of  quantum control   has proved   an immensely productive area of research, with an ever growing number of   applications ranging from quantum chemistry \cite{Levi15,Olson16,McDonald18}, quantum thermodynamics \cite{Xiaoting11,Kosloff13,KNK15}, quantum metrology \cite{PJ17,PCF18} to quantum computing \cite{VC05}.
A fundamental goal in quantum control  is to provide practical control methods that can utilize available control means and   can be robust to noisy environments. To this end, a variety of quantum control models have been studied and a great many quantum control techniques have been put forward \cite{Altafini12}. These   include quantum optimal control \cite{WG07,Glaser15}, coherent control of dissipative systems \cite{Altafini03,VWC09,Koch16,Jun16}, state engineering via  quantum measurements \cite{AN10,PISR06},     closed-loop   control \cite{Rabitz00,LV01,ZLWJN17,Jun17}, etc. In all these studies, attempts were made to establish mathematical modeling of systems and controls, and formulate quantum control problems formally.


Quantum control is realized through the application
of an external action to the system. Often such an external action is implemented via a suitable tailored coherent control field.
There are fundamental bounds on the coherent transformations of   density operators in Liouville space, associated with the fact that unitary transformations should conserve   eigenvalues. 
With non-unitary operations, however, other regions of operators in Liouville space are reachable. Incoherent form of action on the system could be induced in a number of ways, e.g.,  reservoir engineering \cite{SW10,PCZ96} or measurement-driven quantum evolution \cite{Roa06,PISR06}. It is then interesting to ask what  the bounds of state transfer  would  be like  when a certain incoherent control strategy is chosen.  

To address the above  reachability problem, it is desirable to use the mathematical machinery of quantum channels.  
Quantum channels capture in most general terms the input-output relations of quantum devices, covering     various kinds of effects   \cite{MZ12}.  They are used to    describe a broad class   of transformations that a quantum   system can undergo. A channel $\mathcal{\hat E}$ is, by definition,   a completely positive trace-preserving map, acting on a physical system at state $\rho$ to get a new state $\mathcal{\hat E}(\rho)$.   Let $\mathcal{S}$ denote a set of quantum channels composed of a certain type of transformations. Some common examples of   $\mathcal{S}$ are unitary channels, mixed unitary channels and unital channels.    Now the problem of determining bounds of state transfer under $\mathcal{S}$ can be stated formally as such: for an initial state $\rho$ and   a target traceless Hermitian operator $\sigma$, to    optimize 
\begin{equation}
\max_{\mathcal{\hat E} \in \mathcal{S}} D\left(\mathcal{\hat E}(\rho), \sigma \right) = \frac{\operatorname{Tr}\left(\mathcal{\hat E}(\rho)  \sigma \right)}{\operatorname{Tr}\left(  \sigma^2 \right)},
\end{equation}   
where the function $D$     measures how much component of the desired operator $\sigma$ is there in $\mathcal{\hat E}(\rho)$ that has been attained.
Bound of state transfer under unitary channels was established    in Refs. \cite{S90, S95} as the  known \emph{universal bound  on spin dynamics}. The theory then proved to be of great assistance in   the design of experimental control sequences \cite{NHS95,NTS96,HW97,UN00,  KLG03, L16}.
Recently,   bounds on functions of quantum states  under more general quantum channels have also been    analyzed  theoretically \cite{LPW16}.

Following this line of research, the   present paper aims to applying   the   general theoretical framework  of    optimal bounds of state transfer under quantum channels to nuclear magnetic resonance (NMR) quantum state engineering experiments.  Our study is centered around the question that, what is the maximum efficiency for the transfer
between two spin states and how is this accomplished using experimentally available control
fields. In  NMR,  besides of coherent radio-frequency control,   incoherent effects are usually induced through   two common techniques, namely gradient field   and phase cycling. We model the latter two   as mixed unitary channels. In so doing, we then draw a conclusion that, gradient field control and phase cycling can never enhance transfer efficiency, due to that    mixed unitary channels   can   not extend the convertible region beyond coherent controls.   We show that this property facilitates optimal experimental design for state transfer under mixed unitary channels.
On the other hand, it may be possible that coherently controlled relaxation channels  can  surpass  the unitary bounds. To see this, we  experimentally investigate  the problem of  pseudopure state  (PPS) preparation on a two-spin NMR system \cite{CPH, GC, KCL}. To create PPS with polarization as high as possible is of importance for quantum computing tasks. We  compare  experimental results obtained by relaxation-assisted preparation methods, including periodic control proposed in Ref. \cite{Jun16} and line-selective saturation proposed here, with that by conventional methods that employ gradient field or phase cycling, and   find that higher magnitude of PPS can be reached  if relaxation effects are  properly utilized.   We also perform   labelled-PPS preparation experiments to
demonstrate our optimal state transfer strategy under mixed unitary channels.

The organization of the paper is as follows. In Sec. \ref{Framework}, we describe the general   framework of theoretical bounds on state transfer under quantum channels. In Sec. \ref{Model}, we   present quantum control models that are routine in NMR control system. Of particular concerns are the characterization of different kinds of available control means. To be concrete, coherent, incoherent and learning control models are established, thus setting a foundation for a thorough investigation of NMR state preparation problem. Sec. \ref{Experiment} is devoted to   various PPS preparing methods. Experimental results on a concrete two-spin system are given for  making control performance comparisons among the preparation methods. 
In Sec. \ref{Label}, we apply the optimal bound theory to the task of labelled-PPS preparation, and  experimentally demonstrate a concrete example with   a three-spin NMR system. Last, conclusions and prospective views are given in Sec. \ref{Conclusion}.

\section{Theoretical Framework}
\label{Framework}
 
In a   quantum control experiment, one has to consider   two important questions, that is, analysis of controllability  and synthesis of controls. When the goal is to realize a certain desired state, it is worthwhile to check first whether   this state is reachable at all, before starting to search for controls that lead to the target. A physically reasonable transform of a quantum state is normally modeled   as a quantum channel $\mathcal{\hat E}$. Most generally,   $\mathcal{\hat E}$ admits an operator-sum representation, but here  we  restrict our considerations on  the class of   unitary channels, mixed unitary channels and Markovian  channels. In the following, we set up the   necessary theoretical machinery to tackle the problem of optimal bounds on state
transfer.

\subsection{Optimal bounds under unitary channels}

Consider    state transfer via unitary engineering in an $N$-dimensional state space. Since a unitary operation preserves the identity operator, thus it suffices to  consider  the transfer between the traceless part of the states. Now, we use the traceless Hermitian operator $\rho$ to refer to the initial state. The goal is that the final state contains as much of $\sigma$ as possible. The transfer can be described by a unitary transformation $U$ giving
\begin{equation}
\label{unitary}
U \rho U^\dag =   \eta \sigma + \varsigma, \quad \operatorname{Tr}\left( \sigma^\dag \varsigma \right) = 0,
\end{equation}
here,  $\varsigma$ is a residual   operator orthogonal to $\sigma$ and vanishes only when $\rho$ and $ \sigma$ have the same eigenvalue spectra. The coefficient $\eta$, referred to as the \emph{state transfer efficiency}, measures the quality of $U$ in  realizing the state transfer.  To exploit its theoretical   limit, S{\o}rensen 
introduced the concept of \emph{universal bound on spin dynamics} \cite{S90, S95}, according to which,
\begin{equation}
\bm{\lambda}^\rho_\uparrow \cdot \bm{\lambda}^\sigma_\downarrow / \bm{\lambda}^\sigma_\downarrow \cdot \bm{\lambda}^\sigma_\downarrow  \le  \eta   \le   \bm{\lambda}^\rho_\downarrow \cdot \bm{\lambda}^\sigma_\downarrow / \bm{\lambda}^\sigma_\downarrow \cdot \bm{\lambda}^\sigma_\downarrow,	
\label{universalbound}
\end{equation}
here  $\bm{\lambda}^\rho_\uparrow$ and $\bm{\lambda}^\rho_\downarrow$ are vectors containing the eigenvalues of the operator $\rho$, arranged in ascending and descending order respectively, and similarly for $\bm{\lambda}^\sigma_\uparrow$ and $\bm{\lambda}^\sigma_\downarrow$.

\begin{figure}[t]
	\includegraphics[width=0.9\linewidth]{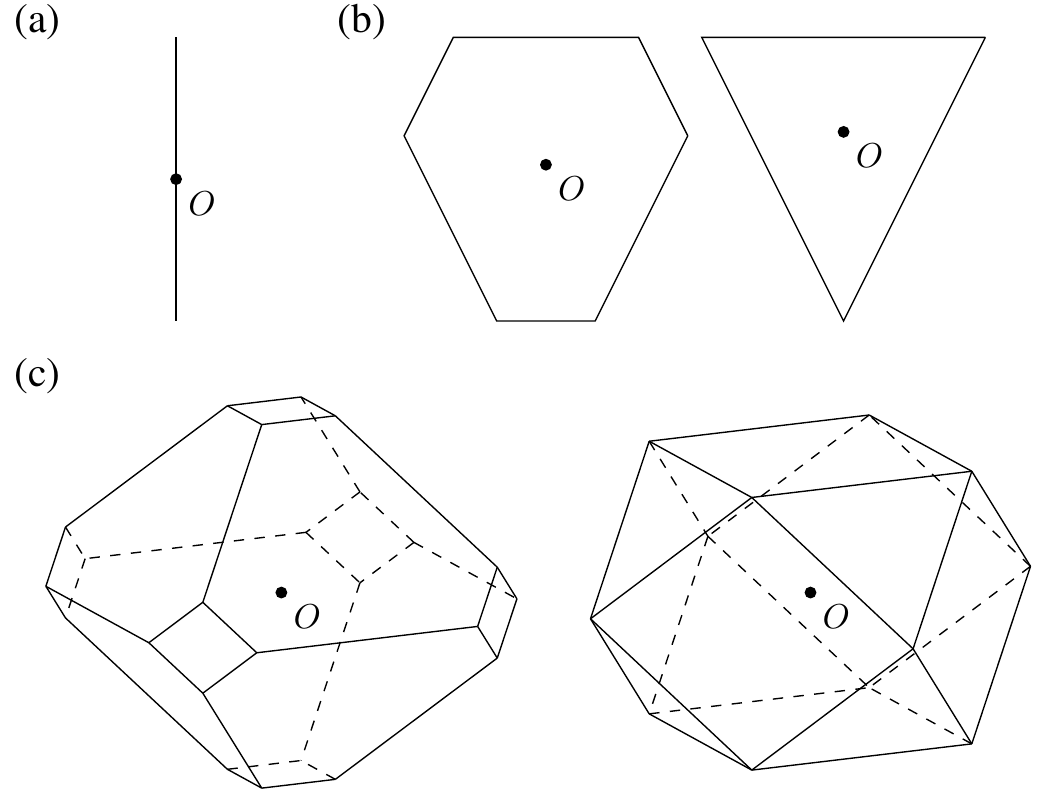}
	\caption{For an $N$-dimensional density operator $\rho$, its universal bound on spin dynamics is geometrically   a permutohedron of order $N$. Here  shows the first examples of permutohedrons:  (a) $N=2$ (line segment); (b) $N=3$ (hexagon); (c) $N=4$ (truncated octahedron). The center point $O$ denotes the maximally mixed state $\bm{1}/N$, here $\bm{1}$ means the identity operator. If some of the eigenvalues  of $\rho$ are equal to each other, then the corresponding permutohedron degenerates.}
	\label{permutohedron}
\end{figure}

The derivation is very neat, which we  briefly present as follows. According to   majorization theory (Ref. \cite{Horn13},  Theorem 4.3.49), the vector of main diagonal entries of $U \rho U^{\dag}$ must be majorized by $\bm{\lambda}^\rho$, and thus can   be written as a convex combination of all permutations of the eigenvalues of $\rho$: $
\bm{d}(U\rho {U^\dag }) = \sum\nolimits_{k = 1}^{{N!}} {{\mu _k}{P_k} \bm{\lambda}^\rho }$,
here $\mu_k \ge 0$, $\sum\nolimits_{k = 1}^{{N!}} {{\mu _k}}  = 1$ and $\left\{ P_k \right\}^{N!}_{k = 1}$ is the group of permutation matrices. Without loss of generality, we assume $\sigma$ to be diagonal. Then, one has
\begin{equation}
\eta   = \frac{\operatorname{Tr}\left( U \rho U^\dag \sigma \right)}{\operatorname{Tr}(\sigma^2)}  
	  = \frac{ \bm{d}(U \rho U^\dag) \cdot \bm{\lambda}^\sigma }{\operatorname{Tr}(\sigma^2)} = \sum\limits_{k = 1}^{{N!}} {\mu _k} \frac{	{ ({P_k} \bm{\lambda}^\rho) \cdot \bm{\lambda}^\sigma}}{\operatorname{Tr}(\sigma^2)}.  \nonumber
\end{equation}
By rearrangement inequality, there is, for any $k$, $\bm{\lambda}^\rho_\uparrow \cdot \bm{\lambda}^\sigma_\downarrow \le ({P_k} \bm{\lambda}^\rho) \cdot \bm{\lambda}^\sigma  \le \bm{\lambda}^\rho_\downarrow \cdot \bm{\lambda}^\sigma_\downarrow$. This gives Eq. (\ref{universalbound}).

These results can be comprehended  from a geometrical point of view.  The set of all spectra majorized by $\bm{\lambda}^\rho$ forms a permutation polytope called permutohedron; see Fig. \ref{permutohedron} for examples. More precisely, let $\bm{\lambda}^\rho = (\lambda^\rho_1, ..., \lambda^\rho_N)$, the permutohedron $\mathcal{P}_N(\bm{\lambda}^\rho)$ is the convex polytope in $\mathbb{R}^N$ defined as the convex hull of all vectors  obtained from $(\lambda^\rho_1, ..., \lambda^\rho_N)$ by permutations of the coordinates. Due to the normalization requirement $\operatorname{Tr}\rho =1$, this polytope lies in the hyperplane $\left\{ (x_1,...,x_N) | x_1 + \cdots + x_N =1 \right\} \subset \mathbb{R}^N$. Thus $\mathcal{P}_N$ has the dimension at most $N-1$. For example,  for $N = 3$ and distinct $\lambda^\rho_1, \lambda^\rho_2,\lambda^\rho_3$, the permutohedron $\mathcal{P}_3(\lambda^\rho_1, \lambda^\rho_2,\lambda^\rho_3)$  is the hexagon; if some of them are equal to each other then the permutohedron degenerates into a triangle, or even a single point.

\begin{figure}[b]
	\includegraphics[width=0.75\linewidth]{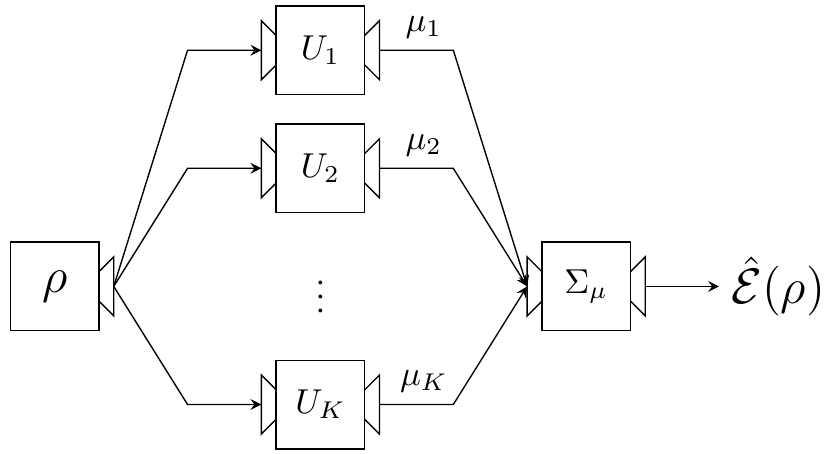}
	\caption{The   statistical average of a series of unitary transform experiments gives a mixed quantum channel $\mathcal{\hat E}$. Here $\mu_k$ is the weight or probability of applying $U_k$.}
	\label{random unitary}
\end{figure}

\subsection{Optimal bounds under mixed unitary channels}
\label{MixedUnitaryChannel}
Now, we turn to   a specific type of non-unitary quantum channel, named as mixed unitary operation, or random unitary operation.  
A mixed unitary channel $\mathcal{\hat E}$ admits a Kraus representation \cite{K83} as
\begin{equation}
\mathcal{\hat E}(\rho) = \sum\limits_{k=1}^K {\mu_k U_k \rho U^\dag_k},
\end{equation}
in which the scalars $\mu_k$ form a probability distribution, i.e., they are non-negative and add up to 1. One physical motivation of studying such channels is the desire to model the effects of classical error
mechanisms present in   the preparation and processing of quantum states. For example,    when a classical parametric uncertainty error   occurs in a state engineering experiment, then
the resulting operation will not be described by a particular unitary, but rather by a mixture of such unitaries; the mathematical description of such a mixture is effectively a random unitary map. 

Mixed unitary channels may also refer  to the scenario where the experiment is designated  as the weighted superposition of several unitary operations, e.g., in the common-used coherence pathway selection method in NMR spectroscopy.
Practically the channel can be realized as such, we perform in total $K$ experiments, for the $k$-th of which we apply the unitary operation $U_k$, and finally add these transformed states; see Fig. \ref{random unitary}.  There has been established the result that  the following two statements are equivalent (Ref. \cite{LP11}, Theorem 3.6): (i) There exists a mixed unitary channel $\mathcal{\hat E}$ such that $\mathcal{\hat E}(\rho) = \sigma$; (ii) the vector of eigenvalues of $\sigma$ is majorized by that of $\rho$. This means that  mixed unitary channels can not extend reachability. The conclusion can also be readily seen as such, since each of the experiments is limited by the universal bound on spin dynamics, their weighted average for sure can not exceed this bound.
Therefore,   for   those mixed unitary channels  that can realize the state transfer $\mathcal{\hat E} (\rho) = \eta \sigma$, the transfer efficiency $\eta$ is bounded by Eq. (\ref{universalbound}). 
Nonetheless, it is to be noted that unlike Eq. (\ref{unitary}),   here   no residual operator appears and  the state transfer is exact.

\begin{figure}[b]
\centering
	\includegraphics[width=0.8\linewidth]{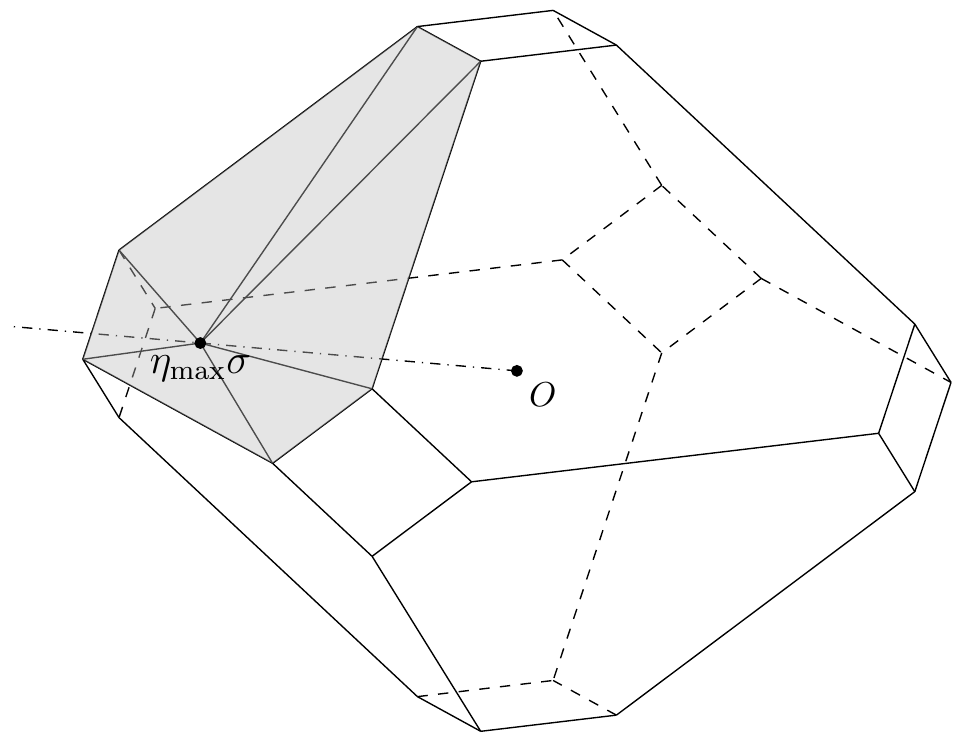}
	\caption{Geometric illustration of how to determine a minimal set of unitaries to prepare a target operator $\sigma$ with optimal   efficiency.}
	\label{design}
\end{figure}

We can further establish  a strategy of optimal experimental  design for state transfer under mixed unitary channels, which achieves best transfer efficiency and meanwhile minimizes the number of unitary experiments required. More precisely, we hope to find a minimal set of unitaries $U_1, ..., U_K$ to optimize
\begin{align}
\min_K \max_{U_1, ..., U_K} & \quad \eta,  \nonumber \\
\text{such that } & \quad \sum\limits_{k=1}^K {\mu_k U_k \rho U^\dag_k} = \eta \sigma.
\label{problem}
\end{align}
It is evident that this problem can be   reduced to the case when     $\rho$ and $\sigma$ are both diagonal. For such a reduced case, we now show that     the solution   is provided by a set of diagonal permutation operations. Let $\mathcal{R}$ denote the set  $\left\{ \eta \sigma: \eta \ge 0 \right\}$, it is a ray starting from the center $\bm{1}/N$. As   the vector of diagonal elements of the operators that are accessible from $\rho$ under   mixed unitary channels is bounded by the permutohedron $\mathcal{P}_N(\bm{\lambda}^\rho)$, so   the  intersection point between $\mathcal{R}$ and $\mathcal{P}_N(\bm{\lambda}^\rho)$  is $\eta_{\max}  \sigma$. This point lies on a certain face of $\mathcal{P}_N(\bm{\lambda}^\rho)$, and suppose      the vertices of that face are $\left\{ P_{s_i} \bm{\lambda}^\rho: {s_i \in S} \right\}$, with $S \subset  \left\{1,...,N! \right\}$ being an index set. Now it is easy to see that, the minimal number of unitaries for preparing $\eta_{\max} \sigma$ equals to   $\operatorname{rank}(\eta_{\max}  \sigma, P_{s_1} \bm{\lambda}^\rho, P_{s_2} \bm{\lambda}^\rho, ...)$, and one can set up a system of linear equations to solve out the necessary permutation operations and the corresponding weights.  This procedure is geometrically illustrated with an example in Fig. \ref{design}. It is to be noted that,  Eq. (\ref{problem}) has many optimal solutions. In practice, to choose  a mixed unitary channel from the optimal ones, needless to say, we  prefer those    diagonal permutation operations that are easier to implement in experiment.

%
%
%

\subsection{Markovian channel under coherent control}

In open quantum system engineering, Markovian channels are often employed to directly describe the underlying noisy physical processes governing the evolution. They are frequently used to model realistic experimental set-ups, especially in   NMR systems under relaxation, quantum optics,  and condensed-matter physics, where external noises must invariably be accounted for. If the   correlation time of the environmental noises is much faster than that of the system, then to a good approximation the underlying physical processes are   forgetful. This is the Markovian assumption. Mathematically,   a Markovian master equation generates a one-parameter (time $t$) semi-group  of CPTP (completely positive and trace preserving) maps.  Under the Markovian assumption, the system evolution under the joint action of coherent control and noise perturbation can be described by the Lindblad equation \cite{Lindblad76,BP02}
\begin{equation}
\dot \rho = \mathcal{\hat L}\rho =-i\left[ {{H_S} + {H_C} ( t ),\rho } \right] + \hat {\mathcal{R}}\rho,
\label{Lindblad}
\end{equation}
where we have set the Planck constant $\hbar$ to be 1, $H_S$ is the system Hamiltonian, $H_C(t)$ is the
time-dependent external control Hamiltonian, and $\hat {\mathcal{R}}$ is a
 superoperator of Lindblad type $\mathcal{\hat R}\rho = \sum_\alpha {\gamma_\alpha (2L_\alpha \rho L^\dag_\alpha - L^\dag_\alpha L_\alpha \rho - \rho L^\dag_\alpha L_\alpha)}$ with $L_\alpha$ being Lindblad operators representing non-unitary effects. 

The Lindblad equation leads to a dynamical map $\mathcal{\hat E}$ which steers an initial state  $\rho(0)$ to another state $\rho(t)$ at time $t$. Because $\mathcal{\hat E}$ is non-unitary, it is possible to make an exact state transfer $\mathcal{\hat E}: \rho \to \eta \sigma$, where $\sigma$ is the target operator. Apparently, the bound for $\eta$   depends on the properties of $\mathcal{\hat R}$. Since decoherence is one of the main obstacles in developing quantum information technology, estimating the impact of decoherence on state transfer efficiency   is hence of great importance to evaluate the realistic performance.  Unfortunately, seeking the limit for reachability that is applicable for a general open Markovian quantum system driven by decoherence process and coherent controls remains an unsolved problem. Only partial results exist. For example, the reachable set on the states of a single qubit under dissipation has been established in Refs. \cite{Yuan10,Yuan12}. Also, enormous efforts have been made to generalize these results to higher dimensional systems  \cite{Altafini03,Altafini04,KDH12,Rooney12,Jun16,Rooney18}. 

Markovian channel under coherent control is interesting for the following reasons. First, a controlled relaxing system leads to a strictly
contractive quantum channel, meaning that for any   fixed pair
of initial states, their distance    is a strictly decreasing function of time \cite{RH12}. Therefore,  it is possible to develop periodic control schemes which promise the capability of driving the system asymptotically to some periodic steady state.  Such a property ensures that the target state can be periodically retained. The other reason, somehow surprisingly, is that open system control may exhibit   advantages  in some circumstances, e.g., in heat-bath algorithmic cooling protocol, higher purification efficiency can be achieved by utilizing relaxation \cite{Ryan08}. Therefore, it is  worth  studying the conditions on which    Markovian channels  could offer enhancement of state transfer efficiency beyond the universal bound on spin dynamics.
In our subsequent experimental part, we will demonstrate these merits with concrete examples.

\section{NMR Based Control Models}
\label{Model}


For a given physical system, recognizing the available control means and establishing proper control models are crucial for developing practical control methods. A control model involves three necessary ingredients: system dynamics, admissible controls and control performance. This section gives a brief description of the relevant control models that are useful    in liquid NMR experiments.

\emph{Coherent control.} There have been developed substantial control techniques in NMR spin systems; for a survey, see Ref. \cite{VC05}.  The most frequently used control technique in NMR spectroscopy is to excite the nuclear spins with radio-frequency (RF) pulses. Let the  natural Hamiltonian describing the coupled spins be denoted by $H_S$, let the time-dependent control Hamiltonian representing RF fields be denoted by $H_C(t)$, then the system state evolves according to the Liouville-von-Neumann equation
\begin{equation}
\dot \rho (t) =  - i[{H_S} + {H_C}(t),\rho (t)].
\end{equation}
This is a closed quantum system control model, where RF excitation is coherent and enables universal control.

\emph{Gradient field control.} Incoherent controls  are often implemented by applying gradient fields, which exploit the different sensitivity of coherence orders to magnetic field inhomogeneity. Pulsed magnetic field gradients are mainly used for    destructing the transverse magnetization of the sample, for on occasions only longitudinal magnetization is desired at some step of the evolution pathway.   Here, the longitudinal direction, denoted as the $\hat z$ axis,  is along the large  static magnetic field $B_0$.
A gradient field along $\hat z$  can introduce a dephasing for the coherences associated with different spatial locations along the sample.

Consider the most simple case that,  a constant inhomogeneous field $B'(z) = gz$ in $\hat z$ direction is applied, where $z \in [-L, L]$ with $2L$ being the length of the sample region, and $g$ is the field gradient, so that the total field is $\bm{B}(z)=B_0 {\hat z}+B'(z) {\hat z}$. The system Hamiltonian is then  $H_z = \gamma (B_0+g z)\sigma_z/2$, here $\gamma$ is the gyromagnetic ratio. Under this Hamiltonian,   the $\nu$-th coherence terms $\sigma_\nu$ will evolve into
\begin{align}
 \sigma^p(t) & =  \exp(-iH_zt)\sigma_\nu(0) \exp(iH_zt)  \nonumber \\
 & =  \exp[- i \nu \gamma(B_0 + gz)t]\sigma_\nu(0).  \nonumber
\end{align}
Taking the ensemble average of the above expression with respect to $z$, there is 
\begin{equation}
\label{gradient}
 \left\langle \sigma_\nu(t) \right\rangle = \int_z {{U_z}(t){\sigma_\nu}(0)U_z^\dag (t)}dz
 =\sigma_\nu(0)\delta_{\nu,0}.
\end{equation}
Therefore, except for zero-th order quantum coherences, all other non-diagonal elements of the density matrix will be eliminated. In other words, one can regard the macroscopic sample as being constituted by a set of sub-ensembles each one represented by a density matrix with the same distribution of populations, but with off-diagonal elements out of phase. Averaging over these sub-ensemble is then equivalent to that the system undergoes a mixed unitary channel.

\begin{figure}[t]
\includegraphics[width=0.75\linewidth]{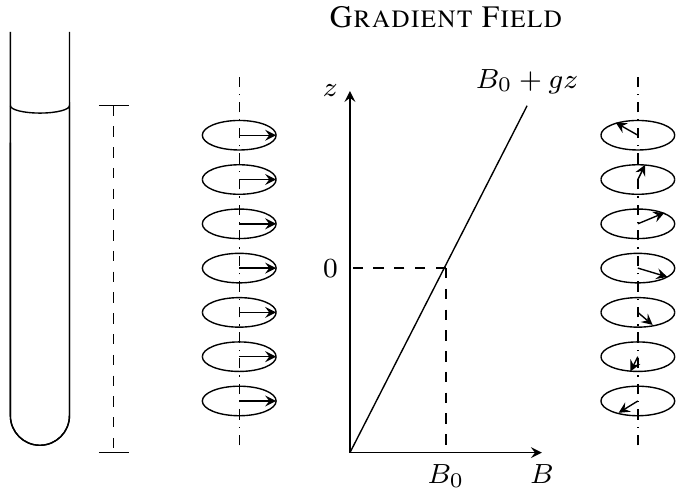}
\caption{Illustration of the non-unitary evolution under   gradient field control. Because the dephasing is proportional to the gradient strength,  the coherences vary from 0 to $2\pi$ along the sample.}
\end{figure}

\emph{Phase cycling.} An alternative to gradient field control  that can be used in many instances to discriminate against particular coherence pathways is phase cycling. 
In many pulse NMR techniques, some sort of phase cycling procedure serves as an integral part of the experiment, e.g., NOESY \cite{Levitt2008}. Phase cycling can be used   for the reduction or even elimination of unwanted signal components, giving rise to signal distortions. It is a method that lies at the very heart of almost every NMR experiment,   especially playing a critical role in 2D NMR. 

Phase cycling refers to the repetition of a pulse sequence a number of times, with varying the phases of the RF pulses, the receiver, and digitizer in a predetermined pattern. With judiciously changing the phases from experiment to experiment,  the final signals are averaged    such that   the desired component is retained while undesired routes are rejected.
To illustrate the idea, consider a simple phase cycling scheme aiming to select wanted coherences. In the $k$th experiment, we apply a collective rotation of the spins with angle $\theta_k$ about the $\hat z$ axis, then in the final averaged signal, any  $\nu$ coherence term $\sigma_\nu$ will be turned into 
\begin{equation}
{\sigma_\nu} \to  {\sigma_\nu}  \sum\limits_k {{e^{ - i{\nu \theta_k}}}}.
\end{equation}
Through a special design of the phases $\theta_k$, the desired coherence terms can be enhanced, 
while the undesired ones are canceled by intendedly setting $\sum\nolimits_k {{e^{ - i \nu \theta _k}}} =0$, when adding up the signals. Generally, phase cycling schemes can be rather complex. It is easy to see that the influence of a phase cycling to quantum states can be directly represented as a mixed unitary channel. Therefore, the phase cycling technique can not exceed the universal bound on spin dynamics.  

\emph{Control of relaxation process.} In NMR, relaxation is always present, and affects the system's state in an irreversible way. The formalism of liquid NMR relaxation theory has been well developed. We refer the reader to standard textbooks  such as Refs. \cite{Levitt2008,KM18} for theoretical details. Under appropriate assumptions, the relaxation dynamics of a liquid-state NMR spin system can be described by the  Lindblad equation (\ref{Lindblad}). A detailed analysis of the state transfer problem under coherently controlled relaxation equation   was given in Ref. \cite{Jun16}, where it was found that when relaxation is properly utilized, it is possible to  surpass the universal bound on spin dynamics. Whether or not this possibility can happen is dependent on the  relaxation parameters in the evolution equation.

\section{Applications to PPS Preparation}
\label{Experiment}

In this section, we   investigate the application of quantum control models to  state engineering tasks in liquid-state NMR quantum information processing.  
Specifically, we focus on the problem of   pseudopure state  (PPS) preparation \cite{CPH, GC, KCL}. Pseudopure states are used as the initial states for quantum computing. From these states, one can prepare other quantum states such as Bell states via quantum circuits.

\subsection{PPS preparation problem}
One important motivation of studying the PPS preparation problem is that, the low polarization of NMR spin ensemble at room temperature makes it  rather difficult  to get a genuine pure state. Therefore, PPS  was introduced as a substitute. It offers a faithful representation for the transformations of pure states. Especially, it turns out that the concept of PPS is suitable for benchmark experiments. 

We consider creating pseudopure states on a spin system with weakly coupled $n$ spin-1/2 nuclei. We want to prepare the following form of mixed state
\begin{equation}
\rho_\text{pps} = (1-\eta) \frac{I^{\otimes n}}{2^n} + \eta \left| 0 ^{\otimes n}\right\rangle \left\langle 0 ^{\otimes n}\right|,
\label{ppsform}
\end{equation}
where  $I$ is the $2  \times 2 $ identity matrix, and $\eta$ is referred to as the \emph{effective purity} of the state. Let $X,Y,Z$ denote the three Pauli operators $\sigma_x, \sigma_y, \sigma_z$ respectively, then PPS can also be written as
\begin{align}
{\rho_\text{pps}} = & {} \frac{I^{\otimes n}}{2^n} + \frac{\eta}{2^n} (Z_1 +  \ldots  + Z_n + Z_1Z_2 +  \ldots + \nonumber \\
& {}  Z_nZ_1 +  \ldots  + Z_1Z_2 \cdots Z_n).
\end{align}
Effective purity is an important parameter since it determines the intensity of the   observed signals. NMR suffers from inherently low sensitivity, and it is desirable to improve the spectral signal-to-noise ratio as much as possible.

The initialization process usually starts with the system's thermal equilibrium state $\rho_{eq}$, which satisfies the Boltzmann distribution and takes the following form if high temperature approximation is used
\begin{equation}
{\rho_\text{eq}} - \frac{I^{\otimes n}}{2^n}  \propto  \epsilon \sum\limits_{i = 1}^n {{\gamma _i}{Z_i}},
\end{equation}
where $\gamma_i$ is the gyromagnetic ratio of the $i$-th nucleus and the factor $\epsilon \simeq {10^{-5}}$ is related to the thermal polarization of the
system. The polarization is actually rather low, which is responsible for the difficulty of producing pure states.

\begin{figure}[b]
	\includegraphics[width=0.9\linewidth]{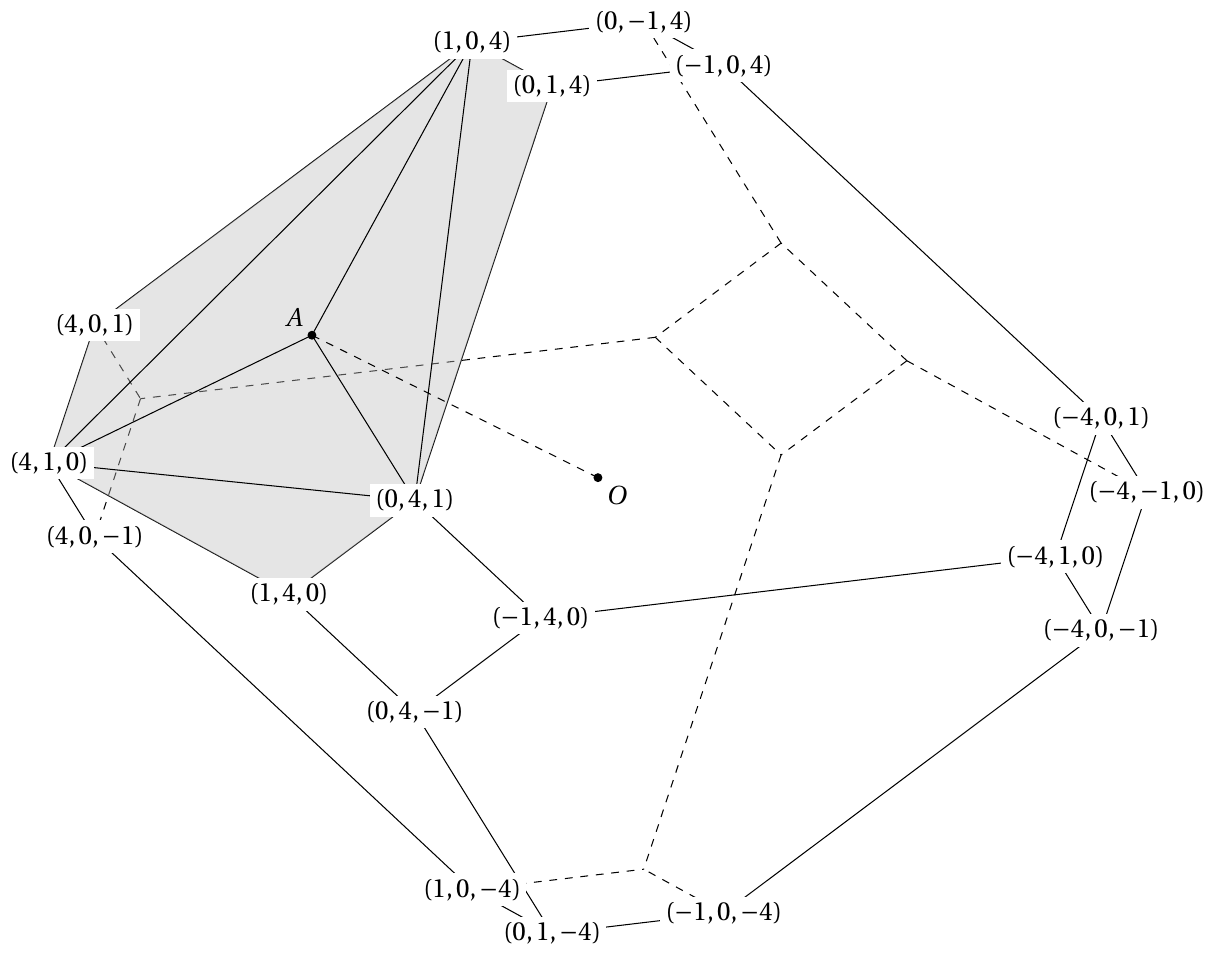}
	\caption{The permutohedron   constructed by taking the convex hull of all 4-vectors from the $4!=24$ diagonal permutations of $\rho_\text{eq}$.}
	\label{convex polytope}
\end{figure}

The PPS task, i.e., $\rho_\text{eq} \to  \rho_\text{pps}$, can not be done merely with unitary operations, as $\rho_\text{eq}$ and $\rho_\text{pps}$ have different spectra. Methods of PPS preparation often involve different ways of realizing non-unitary operations \cite{KLMT00,SOF00,J01},  such as exertion of gradient fields, phase cycling or even relaxation channels. 
To simplify the discussion of the PPS preparation problem, we use a heteronuclear system ($^1$H and $^{13}$C) 
as an example. In spite of this, note that in principle the approaches presented in the following can be  generalized to any number  of qubits. The diagonal part of the state of our two-spin system  can always be decomposed as $II/4 + x_1 Z_1 + x_2 Z_2 + x_3 Z_1 Z_2$, which we denote as a vector $(x_1, x_2, x_3)$. For example, let   $\epsilon$ denote the polarization magnitude of the $^{13}$C spin,  then the equilibrium state  $\rho_\text{eq}$ is written as $(\epsilon,4\epsilon,0)$. The diagonal permutations of $\rho_\text{eq}$ are all shown in Fig. \ref{convex polytope}.
In the language of polarization transfer, our task
amounts to the conversion between the operators
$\rho = \epsilon (Z_1 + 4 Z_2)$ and $\sigma = -II/4 + \left| 00\right\rangle \left\langle 00\right|$.   As discussed in Sec. \ref{Framework}, the universal bound on spin dynamics suggests that the highest unitary transfer efficiency achievable  is
\begin{equation}
	\eta_{\max} =   \frac{\bm{\lambda}^\rho_\downarrow \cdot \bm{\lambda}^\sigma_\downarrow}{\bm{\lambda}^\sigma_\downarrow \cdot \bm{\lambda}^\sigma_\downarrow} = \frac{20}{3} \epsilon,
	\label{eta_max}
\end{equation}  
no matter we use  techniques of gradient fields or phase cycling or both. The unitary bound or mixed-unitary bound is   geometrically illustrated  in Fig. \ref{convex polytope}.

\subsection{Methods for PPS preparation}

Here we present some examples of methods for PPS preparation, with major concern on their transfer efficiencies. In the following, method (1) is an average of multiple unitary control experiments; methods (2) spatial averaging method, (3) line selective method, and (4) controlled-transfer gate method  all utilize gradient fields to realize necessary non-unitary controls.  They can achieve transfer efficiency at most $\eta_{\max}$ in Eq. (\ref{eta_max}). We also describe two novel PPS preparation methods (4) and (5), which take the relaxation effects as  the required non-unitary resource  and remarkably, both allow   to surpass the limit $\eta_{\max}$. We use a concrete two-spin system as a simple application example to show the transfer efficiencies for these preparation methods. The system is $^{13}$C-labeled chloroform $ ^{13}$CHCl$_3$, which consists of a carbon spin and a proton spin. We use $J$ to denote the coupling between the two spins.

(1) \emph{Time averaging (TA) method. }  This method  \cite{KCL} simply  involves averaging over three unitary experiments, i.e.,
\begin{equation*}
\begin{rcases}
  \rho_\text{eq} \to Z_1 + 4 Z_2  & \\
  \rho_\text{eq}   \to Z_2 + 4 Z_1Z_2 &  \\
  \rho_\text{eq}  \to 4Z_1 +  Z_1Z_2 &  
\end{rcases}
\to 
\rho_\text{pps} = \frac{5}{3}  (Z_1 + Z_2 + Z_1Z_2).
\end{equation*}
From Fig. \ref{convex polytope}, we know that at least three   unitary experiments are needed to get PPS, and there actually exist many choices for these unitaries to achieve the optimal bound $6.67\epsilon$.

(2) \emph{Spatial averaging (SA) method.}
SA \cite{CPH} exploits a series of RF pulses sequentially and repeated field gradient pulses, equalizing gradually the populations of all energy levels except one.

For the CH-two-spin system, we apply the pulse sequence   $[R_x^{\text{C,H}}(90^\circ )  -  1/(4J) - R_y^{\text{C,H}}(90^\circ ) - 1/(4J) - R_{ - x}^{\text{C,H}}(90^\circ ) - \operatorname{G}_z - R_x^{\text{C,H}}(45^\circ ) - 1/(2J) - R_{ - y}^{\text{C,H}}(30^\circ ) - \operatorname{G}_z]$,
where $R_{\pm x (\pm y)}^{\text{C,H}}(\theta)$ denotes a hard pulse rotating the $^1 \text{H}$ and $^{13}\text{C}$ nuclei about $\pm x (\pm y)$ axis with $\theta$ degree and $\operatorname{G}_z$ denotes a gradient pulse along $\hat{z}$ direction to eliminate off-diagonal elements of the state. This transforms the operator $ZI + 4 IZ$  to $5\sqrt{3}/\sqrt{2} (- II/4 +  \left| 00\right\rangle \left\langle 00\right|)$.
Therefore, the transfer efficiency is $\eta = 5\sqrt{3}/\sqrt{2}\epsilon \approx 6.12\epsilon$.
SA  does not achieve the  universal   bound on spin dynamics, since there involves  in a number of gradient pulses, each one causing polarization lost and hence reducing the final spectral signal.

(3) \emph{Line selective (LS) method.}  In this method, one first coherently drive the system from $\rho_\text{eq}$ to a state satisfying that the diagonal population distribution  of which takes the same form as that of $\rho_\text{pps}$, and then applies a  gradient pulse to remove the off-diagonal elements while  keeping the diagonal elements  unchanged.  Line-selective pulses  are designed  to implement the wanted population flips between the target energy levels. By leaving the ground state untouched and averaging the other states,   LS method can acquire the general bound of spin dynamics \cite{MBE98, P01, Zheng18}. 

Generally, the LS method requires to find a set of parameters $\{ {x_\alpha }\} _{\alpha  = {1,\ldots,2^{n-1}}}$, so that the following   unitary operator $U_d =   e^{ -i  \sum\nolimits_\alpha {  {x_\alpha }I_x^{(\alpha  + 1,\alpha  + 2)}} }$, where $I_x^{(\alpha  + 1,\alpha  + 2)}$ is the single quantum transition operator
between levels $\alpha + 1$ and $\alpha + 2$, can fulfill   the  equation
\begin{align}
{} &	 \operatorname{diag}\left[{U_d}\left({\rho _\text{eq}} - \frac{{{I^{ \otimes n}}}}{{{2^n}}} \right)U_d^\dag \right]   \nonumber \\
= {} & 	{\eta _{\max }} \times \operatorname{diag} \left[ - \frac{{{I^{ \otimes n}}}}{{{2^n}}} + \left| {{0^{ \otimes n}}} \right\rangle \left\langle {{0^{ \otimes n}}} \right| \right]. \nonumber
\end{align}
For example, in the C,H-two-spin system, the  operator 
\[{U_d} = \exp \left\{ { - i\left( {{x_1}I_x^{(2,3)} + {x_2}I_x^{(3,4)}} \right)} \right\},\]
with ${x_1} = {31.78^\circ }$ and ${x_2} = {46.50^\circ }$, can be used to create PPS with the highest unitary transfer efficiency $\eta \approx 6.67\epsilon$.
Here
$I_x^{(2,3)} = IX - ZX$ and $I_x^{(3,4)} = XI - XZ$ are quantum transition operators related to the transitions $\left| {10} \right\rangle  \leftrightarrow \left| {11} \right\rangle $ and $\left| {01} \right\rangle  \leftrightarrow \left| {11} \right\rangle $, respectively. For more complex systems, the parameters $\{ {x_\alpha }\} $ can be obtained by numerical searching algorithms.

\begin{figure}[t]
\includegraphics[width=0.4\linewidth]{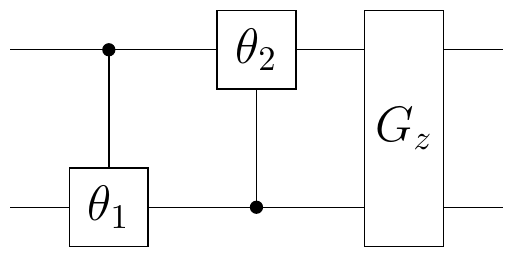}
\caption{The pulse sequence for two-qubit PPS preparation with controlled-transfer gate.
}
\label{CTM}
\end{figure}

(4) \emph{Controlled-transfer gate (CTG) method.}
Similar to  the LS method, the CTG method  allows maximal unitary transfer efficiency also via the idea of averaging  the populations of states other than the ground state. It utilizes the so called controlled-transfer gate, which includes two controlled-rotation gates and one gradient field, to realize the redistribution process (Fig. \ref{CTM}) \cite{Kawamura2010}. The controlled-rotation axes are in the $xy$ plane and the effectiveness of the method does not depend on the specific choice
of the axes. Rotation angles should be specially designed according to the initial  populations of the state. Set rotation angles as $\pi/2$ and repeat the CTGs several times with long enough intervals, the state can be asymptotically driven toward a PPS whatever the initial state is.
For the C,H-two-spin system, one CTG is enough to achieve 
PPS transfer efficiency $\eta \approx 6.67\epsilon$ from the thermal equilibrium state by choosing ${\theta _1} = 99.59^\circ$ and ${\theta _2} = 90^\circ$.

\begin{figure*} 
\centering
\includegraphics[width=0.95\textwidth]{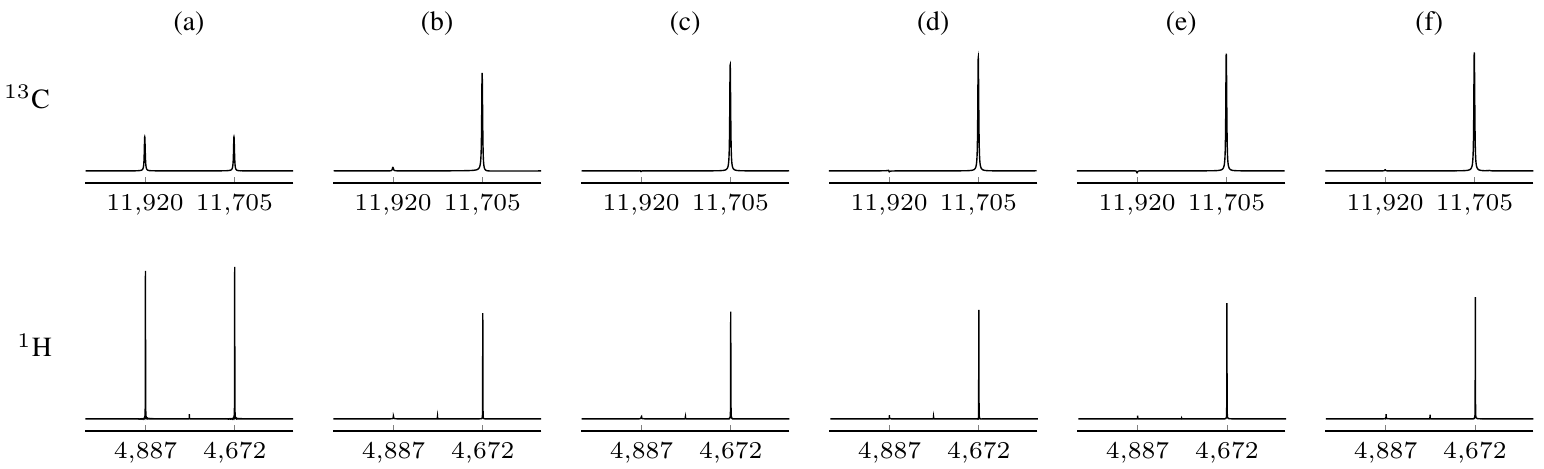}
\caption{Experimental demonstration of pseudopure states prepared using previous approaches and our new approaches. From left to right, (a-g) are spectra obtained by applying readout pulses to the thermal equilibrium state (a), PPS prepared by spatial averaging method (b), by line selective method (c), by controlled-transfer gate method (d), by periodic control (e), and by line selective saturation method (f), respectively. The small signal visible at the center of the 1H spectrum arises from an impurity (unlabeled formate). All spectra are plotted on the same vertical scale.}
\label{PPSspectra}
\end{figure*}

(5) \emph{Periodic control (PC) method.}
In Ref. \cite{Jun16} it was shown that  a simple periodic sequence described by ${\left[ {V  -  \tau } \right]_m}$ is capable of preparing   PPS, where $V$ is a unitary operation, $\tau$ represents  free relaxation, and $m$ means the number of times that the sequence is repeated. Under this sequence,   the total evolution superoperator is $\mathcal{\hat E}  = {(\mathcal{\hat E}_V \mathcal{\hat E}_\tau)^m}$. Our goal is to construct such a sequence so that the desired state $\rho_\text{pps}$ is the corresponding fixed point: $\mathcal{\hat E} \rho_\text{pps} = \rho_\text{pps}$.

During the free evolution periods, the NMR system is under both longitudinal and transverse relaxation, with their respective characteristic time denoted by $ {T_1}$ and ${T_2}$. The longitudinal relaxation tends to steer the diagonal terms of the   state   to the equilibrium distribution, and the transverse relaxation destructs all non-diagonal terms. It is straightforward to conceive a cyclic permutation process of the diagonal terms to realize PPS preparation. For example, $V$ can be chosen to be
\[V = \left[ {\begin{array}{*{20}{c}}
	1 & {} &   {} & {} &   {}  \\
	{} & {} &   {} & {} &   1  \\
	{} & 1 &   {} & {} &   {}  \\
	{} & {} &    \ddots & {} &   {}  \\
	{} & {} &   {} & 1 &   {}  \\
	\end{array}} \right].\]
It is easy to see that $V$ is to perform a permutation of the diagonal elements except the first element ${\rho _{11}}$, and it satisfies ${V^{({2^n} - 1)}} = \text{identity}$. If no relaxation is involved, the permutation operation does nothing but changing the order of the diagonal elements. However, if there are relaxation processes, relaxation would slowly and continuously change the diagonal elements (except $\rho_{11}$) until all of them are equal. In other words,   PPS is a fixed point of $V$ and becomes achievable due to the non-unitary modulation by relaxation.
Note that, unlike in LS method and CTG method, the introducing of a longitudinal relaxation process   may increase $\rho_{11}$, while   coherent controls   can not. Therefore, PC method provides us a route to surpass the highest transfer efficiency in previous methods.

For the C,H-two-spin system, $V$ can be easily decomposed into elementary gates as $[R^\text{H}_y(90^\circ) - 1/(2J) - R^\text{H}_x(90^\circ) - R^\text{C}_y(90^\circ) - 1/(2J) - R^\text{C}_x(90^\circ)]$.

(6) \emph{Line selective saturation (LSS) method.}
It is a common NMR technique to average two energy populations by continuously exposing them to a resonant RF wave. As an extension of this technique, here we put forward a new approach to prepare PPS via saturating the involved energy populations. A shaped RF wave is tailored to simultaneously excite energy level inversions between any two energy states excluding the ground state. In the case that NMR spectral lines can be well resolved, low power square waves of long durations can accomplish   this task. The irradiation should keep on until the saturation condition is achieved. The time scale of this process is often tens of seconds, longer than the longitudinal relaxation time $T_1$. During the irradiation, the generated off-diagonal terms are gradually eliminated by   transverse relaxation. Similarly with the PC method, the longitudinal relaxation may lead to a growth of ground state population  $\rho_{11}$ due to cross-relaxation. This   makes it possible to surpass the bound of the unitary transfer efficiency.

\begin{table}[b]
\centering
{\renewcommand{\arraystretch}{1.5}\setlength{\tabcolsep}{6pt}
\begin{tabular}{|c|c|c|c|c|c|}
\hline
Method & SA & LS & CTG & PC & LSS\\
\hline
Transfer efficiency & $5.60\epsilon$ & $6.21\epsilon$ & $6.60\epsilon$ & $6.83\epsilon$ & $6.88\epsilon$\\
\hline
\end{tabular}}
\caption{Experimental comparison of  transfer efficiencies among different PPS approaches.} 
\label{Comparison}
\end{table}

\subsection{Experimental results}

Our experiments were carried out  on a Bruker AVANCE spectrometer  with a nominal ${}^1$H frequency of 600 MHz at room temperature. The experimental system is the two-qubit heteronuclear spin system (${}^{1}$H and ${}^{13}$C)   provided by  ${}^{13}$C enriched chloroform ($\rm CHC{l_3}$) dissolved in deuterated acetone. ${}^{13}$C nucleus is labeled as qubit 1 and ${}^{1}$H nucleus is labeled as qubit 2. Both nuclear spins were placed on resonance in their respective rotating frames, so the effective
system Hamiltonian contains only a scalar coupling term $H_S = \pi J Z_1 Z_2/2$ with $J=214.5$ Hz. 

We have realized PPS preparation from the aforementioned different methods.
The creation of PPS requires both coherent and incoherent controls. Incoherent controls were implemented   either by applying  gradient field pulses of 1 ms duration or via free relaxation evolution. 
For the SA method, we only need     hard pulse rotations as well as  free evolution in between.
For the LS method and CTG method, we engineered the required unitary operations as   shaped pulses which are optimized by   the gradient ascent pulse engineering (GRAPE) algorithm \cite{Khaneja2005}. The  shaped pulses were functionalized to equalize the populations of all energy levels except the first one, namely to leave the ground state untouched. For the PC method, the period $\tau$ of free relaxation evolution  of each cycle was set to $300$ ms. Then PPS was reached after $m = 60$ loops. For the LSS method,
the line selective saturation operation  was realized by two low power rectangular pulses, whose frequencies are set to be resonant with the energy level transitions  
$\left| {10} \right\rangle  \leftrightarrow \left| {11} \right\rangle $ and $\left| {01} \right\rangle  \leftrightarrow \left| {11} \right\rangle $, respectively. One can also choose to use  pulse shapes other than a rectangle one, such as  a selective soft Gaussian shape.  The two excitation pulses were simultaneously applied, one to the ${}^{13}$C channel and the other to the ${}^{1}$H channel. After continuous irradiation over 3 seconds,  the  populations of the   three energy eigenlevels $\left| {10} \right\rangle$, $\left| {01} \right\rangle$, and $\left| {11} \right\rangle$  were averaged, and then we got a PPS.

  
Figure \ref{PPSspectra} shows the experimental NMR spectra and Tab. \ref{Comparison} shows the comparison between the experimental transfer efficiencies achieved from different PPS approaches. One can see that relaxation assisted methods can outperform mixed unitary channels. The deviations between theoretical expectations and experimental results mainly arise from the imperfections of RF pulses.

\section{Application to Labelled-PPS Preparation}
\label{Label}

\subsection{Labelled-PPS}
Since that in many circumstances PPS is still challenging to prepare, one uses labelled-PPS as a substitute,  which takes the form:
\begin{equation}
	\rho_\text{lpps} = Z \otimes |0 \rangle \langle 0 |^{\otimes (n-1)}.
\end{equation} 
Thus in an $n$-qubit labelled-PPS,  there are $n-1$ qubits being  initialized at $|0 \rangle \langle 0 |$. Preparing labelled PPS can be less resource demanding  than preparing PPS. Indeed, this has been experimentally realized in a 3-qubit system \cite{KLMT00} and later even in a 12-qubit system \cite{Negrevergne06}. 

\subsection{Conventional method}
The standard method for preparing labelled-PPS employs phase cycling. The process can be summarized as follows
\begin{align}
\rho_\text{eq} & \to \rho_0 = Z {I^{ \otimes (n - 1)}} \nonumber \\
& \to \rho_\text{ad}={X^{ \otimes n}} \nonumber \\
& \to \rho_\text{mc}=\frac{1}{2^n}(X+i Y)^{ \otimes n} + \frac{1}{2^n}(X-i Y)^{ \otimes n} \nonumber \\
& \to \rho_\text{lpps} = Z \otimes |0 \rangle \langle 0 |^{\otimes (n-1)}.
\end{align}
The first step  can be done through gradient fields.
The intermediate state $\rho_\text{ad}$ is composed of all anti-diagonal terms. So the coherence orders contained in  $\rho_\text{ad}$ are $-n,-n+2,\ldots,n$.  The maximal coherence state $\rho_\text{mc}$ is    the deviation density matrix for the cat state, i.e., the $n$-coherence $|{{0^{ \otimes n}}} \rangle \langle {1^{ \otimes n}}|+|{{1^{ \otimes n}}} \rangle \langle {0^{ \otimes n}}|$. In the final step, the maximal coherence is   turned to the labelled-PPS $\rho_\text{lpps}$ via a decoding procedure.

\begin{figure*}[htp]
\centering
\includegraphics[width=0.95\textwidth]{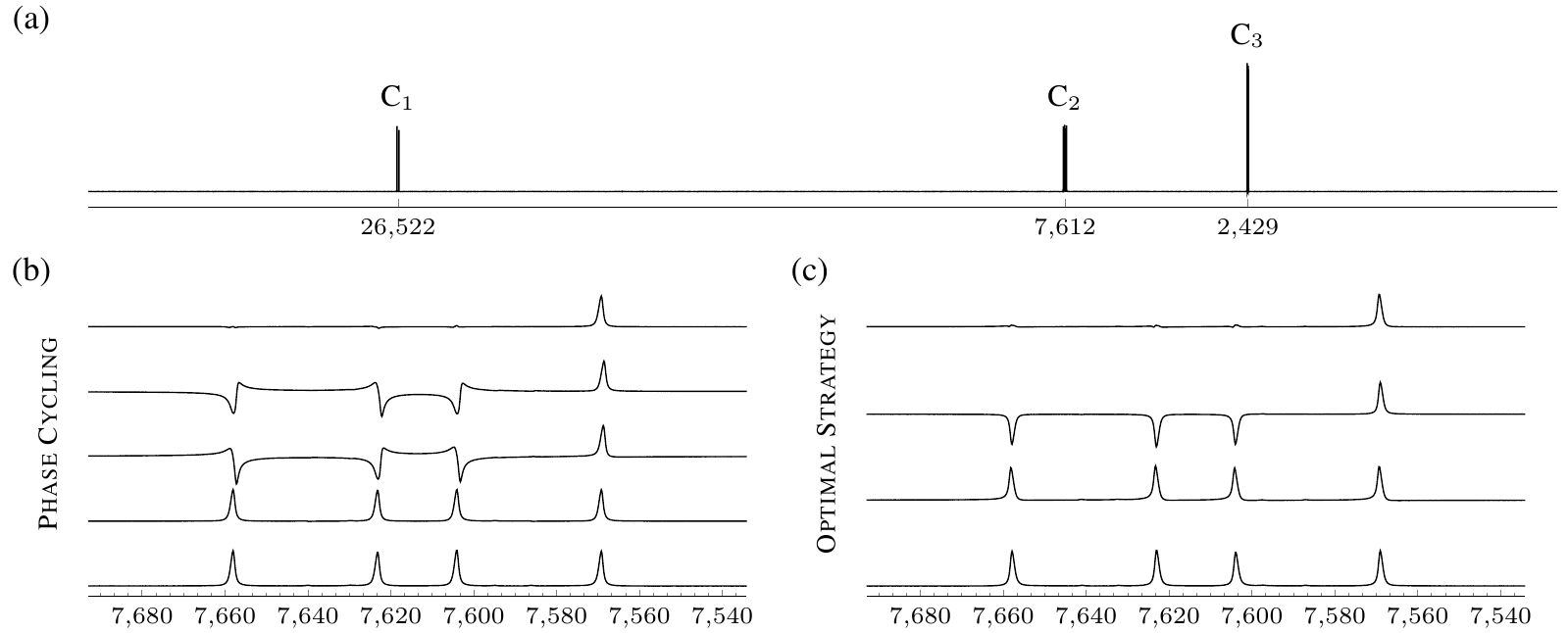}
\caption{Experimental results of labelled-PPS preparation using conventional phase cycling method and our optimal strategy. (a) Full NMR equilibrium spectrum. 
where the chemical shifts (in Hz) of C$_1$, C$_2$ and C$_3$ nuclei are marked. 
(b,c) C$_2$ NMR spectra for labelled-PPS obtained by the phase cycling method (b) and by our optimal strategy (c). The middle three spectra in (b) and two spectra in (c) are obtained from implementing each  unitary operation of the mixed unitary channel described in the main text. The bottom spectra in (b) and (c) are thermal equilibrium spectra of C$_2$ nucleus, and the top ones are the spectra of labelled-PPS, which are obtained by averaging the middle spectra. }
\label{LPPS}
\end{figure*}

The key step of the above method is to make the transform    from $\rho_\text{ad}$ to $\rho_\text{mc}$. To accomplish this task, phase cycling   is adopted to eliminate all the      coherences whose order not equal to $n$ or $-n$. Concretely, we     perform $K$ experiments, where in the $k$th experiment, we   rotate each qubit by a phase $\theta_k$ about the $z$-axis,  that is, we apply the following operation
\[ U_k = \exp (-i \theta_k \sum_{j=1}^n  Z_j/2).\]
Let $\sigma_\nu$ denote   a $\nu$-coherence ($-n \le \nu \le n$), it acquires a phase $\nu\theta_k$ under $U_k$. The  imposed phases thus  label different orders of   coherences. We discriminate two different cases. 

(1) If $n$ is odd,  implying that no zero-coherence exists in $\rho_\text{ad}$, then we   set $K = n$ and $\theta_k = 2\pi k /n$. We average   the results of the $K$ experiments. One can check that, for a $\nu$-coherence $\sigma_\nu$, there is
\begin{numcases}{\sigma_\nu \to \frac{1}{K}\sum_{k=1}^K e^{-i \nu \theta_k}  \sigma_\nu = }
\sigma_\nu, & \text{if } $\nu = \pm n$;  \nonumber \\
0, & \text{otherwise. }  \nonumber
\end{numcases}

(2) if $n$ is even, implying that there exist   zero-coherences in $\rho_\text{ad}$, then we   set $K = 2n$ and $\theta_k =   2\pi k /(2n)$. The results of the $K$ experiments are treated as such,  the $k$th experimental result is added when $k$ is even and subtracted when $k$ is odd
\begin{numcases}{\sigma_\nu \to \frac{1}{K}\sum_{k=1}^K (-1)^k e^{-i \nu \theta_k}  \sigma_\nu = }
\sigma_\nu, & \text{if } $\nu = \pm n$;  \nonumber \\
0, & \text{otherwise. }  \nonumber
\end{numcases}

Therefore, for these two cases, the above designed phase cycling   preserves exclusively the maximal order coherences. The    final signal can only originate from the   $n$-coherences in $\rho_\text{ad}$.
To summarize,  the number of experiments need to be performed in this labelled-PPS preparation method  is $n$ when $n$ is odd, and is $2n$ when $n$ is even.

\subsection{Our optimal strategy}
\label{lpp_optimal}
Here, we show that, actually just two experiments suffice to get the labelled-PPS. This is based on     optimal experimental design   under mixed unitary channels, which has been developed in Sec. \ref{MixedUnitaryChannel}.
One optimal solution is given as below
\begin{equation}
\label{optimal}
\frac{1}{2} \left(Z {I^{ \otimes (n - 1)}} +   U \cdot Z {I^{ \otimes (n - 1)}} \cdot {U^\dag } \right) = Z \otimes | 0\rangle \langle  0|^{ \otimes (n - 1)},  \nonumber
\end{equation}
where
\[ U = \left[{\begin{array}{*{20}{c}}
   {\rm{1}} & {} & {} & {}  \\
   {} & {} & {} & 1  \\
   {} & {} &  {\mathinner{\mkern2mu\raise1pt\hbox{.}\mkern2mu
 \raise4pt\hbox{.}\mkern2mu\raise7pt\hbox{.}\mkern1mu}}  & {}  \\
   {} & 1 & {} & {}  \\
\end{array}} \right].\] 
So in the first experiment, we apply an identity operation, i.e., do nothing.  In the second experiment, we apply the above operation $U$. Then we get labelled-PPS of optimal transfer efficiency.
There exist substantially many optimal preparation strategies, which are essentially equivalent due to symmetry.

\subsection{Experimental results}

We carried out labelled-PPS preparation on a three-qubit system. The system is provided by a ${}^{13}$C-labeled    alanine dissolved in deuterated acetone. The sample consists of three ${}^{13}$C nuclei, which are labelled as C$_1$ (qubit 1), C$_2$ (qubit 2) and C$_3$ (qubit 3), respectively. Figure \ref{LPPS}(a) gives the thermal equilibrium spectrum of the sample. The   mutual couplings between the spins are $J_{12}=53.97$ Hz, $J_{23}=34.9$ Hz and $J_{13}=-1.31$ Hz. Observation is made on C$_2$ because all the couplings are adequately resolved.

\emph{Conventional method}. We start from the state $Z_2$, which can be readily obtained by    first rotating the polarizations of  C$_1$ and C$_3$ into the transverse plane, and then applying a gradient pulse. 
Then, the gate sequence $[R_y^2({90^\circ }) - {e^{ - i\pi {Z_1}{Z_2}/4}} - R_y^{\rm{1}}({90^\circ }) - {e^{ - i\pi {Z_1}{Z_3}/4}} - R_y^{\rm{3}}({90^\circ })]$ can be used to make the transform $Z_2 \to Y_1Y_2X_3$. However, it is hard to find a simple pulse sequence to implement the operation $e^{ - i\pi {Z_1}{Z_3}/4}$ within a duration much shorter than the transverse relaxation time, since the coupling $J_{13}$ is so weak. Therefore the gate sequence is integrated into a shaped pulse of 40 ms in the experiment, which is optimized by using the  GRAPE  technique. Next, we encode phases to the  different  coherences  contained in $Y_1Y_2X_3$. As we have described previously, since the state $Y_1Y_2X_3$ contains just one-coherences and three-coherences,  so three experiments are needed for the phase cycling procedure. The rotation angles of the three experiments are set to $0^\circ$, $120^\circ$ and $240^\circ$ respectively. After the encoding, another GRAPE pulse of 40 ms, corresponding to the sequence
$[R_{ - y}^3({90^\circ }) - {e^{ - i\pi {Z_1}{Z_3}/4}} - R_{ - y}^{\rm{1}}({90^\circ }) - {e^{ - i\pi {Z_1}{Z_2}/4}} - R_x^{{\rm{1,3}}}({90^\circ })]$, was applied to decode the selected three-coherence into labelled-PPS, i.e., the state $|0 \rangle \langle {0}| \otimes X \otimes |0 \rangle \langle {0}|$.


\emph{Our method}. The labelled-PPS can be prepared by a mixed unitary channel consisting of two unitary operations. We use the optimal strategy given in Sec. \ref{lpp_optimal}. Each operation is realized  by a GRAPE pulse of 40 ms. The experimental result  is shown in Fig. \ref{LPPS}, which confirms the validity of our optimal strategy for   labelled-PPS preparation.

\section{conclusions and discussions}
\label{Conclusion}

Quantum state transfer represents one of the most fundamental problems in quantum engineering  \cite{DP10,BCR10} and has a wide range of applications \cite{Zhang05,QWL13,QWCL14,JunOTOC,Chiu18}. Very often, it is important to ask what transfer efficiency can be achieved via available control means. In this work, we have examined this problem in detail.  We show that the bound of unitarily convertible region, namely the optimal bound on spin dynamics,  can not be extended   through introducing mixed unitary channels, which are defined as probabilistic  or weighted average of a set of unitary control experiments. Moreover,  having prior knowledge of the optimal bound  under mixed unitary channels, a minimal set of unitaries can be constructed to achieve the best transfer efficiency. This gives optimal experimental design strategy. We explore the application of   state transfer bound theories to realistic state engineering tasks in liquid NMR systems. Specifically, both  theoretical and experimental investigations are conducted on the PPS preparation problem. We show that larger transfer efficiency can be achieved, with the aid of relaxation effects, which are usually considered as playing   negative roles in quantum system control.

Applying the optimal strategy of designing mixed unitary channels to NMR spectroscopy is   a subject worthy of study. Some   NMR techniques, such as  COSY and INADEQUATE experiments \cite{Levitt2008}, whose cores are phase cycling, may gain  improvements of experimental efficiency if optimal strategy can be constructed. As to the theoretical aspect,
our present study can be regarded as a step in exploring   the bound of state transfer under general quantum channels. Future work will consider the extension of the theory to other non-unitary control models, such as quantum measurement \cite{Buffoni19}, reservoir engineering \cite{VWC09,Vuglar18} and hybrid quantum-classical control \cite{Jun17,Yang17}, which have already attracted great interests in recent quantum information processing  experiments.

\section{Acknowledgments}
W. Z., H. W, J. L., T. X., and D. L. are supported by the  National Natural Science Foundation of China (Grants  No. 11605153, No. 11605005, No. 11875159  and No. U1801661),   Science, Technology and Innovation Commission of Shenzhen Municipality (Grants No. ZDSYS20170303165926217 and No. JCYJ20170412152620376),  Guangdong Innovative and Entrepreneurial Research Team Program (Grant No. 2016ZT06D348), the National Natural Science Foundation of Zhejiang province (Grants No. LQ19A050001).


%

\end{document}